\documentclass[floatfix,twocolumn,prl,longbibliography]{revtex4}
\usepackage{graphicx}
\usepackage{dcolumn}
\usepackage{bm}
\usepackage{amsmath}
\usepackage{times}
\usepackage{color}
\usepackage[breaklinks=true,colorlinks,citecolor=blue,linkcolor=blue,urlcolor=blue]{hyperref}

\makeatletter

\newcommand{\Rmnum}[1]{\expandafter\@slowromancap\romannumeral #1@}
\makeatother

\begin{document}

\title{Metamagnetic quantum criticality in the antiferromagnetic topological insulator MnBi$_2$Te$_4$}
\author{Tithiparna Das}
\affiliation{Department of Physics, Indian Institute of Technology Kanpur, Kanpur 208016, India}
\author{Soumik Mukhopadhyay}
\email{soumikm@iitk.ac.in}
\affiliation{Department of Physics, Indian Institute of Technology Kanpur, Kanpur 208016, India}

\begin{abstract}
Combining transport and magnetization measurements, we discover the emergence of a tricritical point connecting the antiferromagnetic, metamagnetic and paramagnetic regions in MnBi$_2$Te$_4$, a magnetic topological insulator candidate which can potentially exhibit axion electrodynamics. We observe a unique magnetic field-driven phase diagram where the second-order antiferromagnetic to paramagnetic phase boundary bifurcates into two first-order lines at the tricritical point. The two first-order lines enclosing the metamagnetic phase eventually terminate at the quantum critical endpoints with the application of magnetic field.
\end{abstract}

\maketitle

{\it Introduction:--} Topological Insulators (TIs) have insulating bulk and conducting surface states protected by time-reversal symmetry (TRS) \cite{{RevModPhys.82.3045},{RevModPhys.83.1057}, Moore2010, schindler, Weber_2024}. The TRS breaking induced by magnetic order opens up a gap in the topological surface states (TSSs) near the Dirac point \cite{Ji_2021}. The interplay of magnetism and topology in magnetic topological insulators can lead to novel phenomena like quantum anomalous hall effect (QAHE) \cite{Chang, Liu, Krieger, He}, topological magnetoelectric effect, etc. \cite{{Tokura2019},{Liu2020},{Essin}, {Dziom2017}}. Thus the electronic band structure and the nontrivial spin texture of magnetic topological insulators (MTI) are fascinating areas of research nowadays. In this respect, the study of accessibility of the multiplicity of magnetic ground states or in other words, of quantum phase transition in such systems is of paramount importance. The emergence of quantum criticality associated with quantum phase transition (QPT) can lead the system through a point of instability at $\mathrm{T = 0}$, known as the quantum critical point (QCP) which separates two distinct stable phases~\cite{{Sachdev_2011},{Coleman2005}}. For a broad class of 3D antiferromagnets, the magnetic field is a useful tuning parameter which can suppress the Neel temperature ($\mathrm{T_N}$) to a QCP \cite{{13Wu},{Kitagawa}, {14Lester},{15Das},{16Millis},{17Belitz},{18Harrison}, {Moya2022}}. A metamagnetic transition in an antiferromagnet is a first-order transition from a low magnetization to a high magnetization state as an applied magnetic field H is swept through a (temperature dependent) critical value $\mathrm{H_{mm}(T)}$. The trajectory of first order transitions $\mathrm{H_{mm}(T)}$ usually terminates at a critical point $\mathrm{(H^\ast, T^\ast)}$ in the $\mathrm{(H, T)}$ plane. However, by appropriately tuning external non-thermal parameters such as pressure, it is possible to reduce $\mathrm{T^\ast}$ to zero, leading to a quantum-critical endpoint (QCEP) \cite{16Millis}. The observation of quantum critical endpoints without tuning is unusual except in rare cases of itinerant metamagnets such as Sr$_3$Ru$_2$O$_7$ \cite{Kitagawa}. Moreover, in itinerant metamagnets at low temperatures, universal quantum effects can lead to a quantum triple point where the ferromagnetic (FM), antiferromagnetic (AFM), and paramagnetic (PM) phases all coexist in the presence of a magnetic field \cite{ 17Belitz}. 

\begin{figure}
\includegraphics[width=\linewidth]{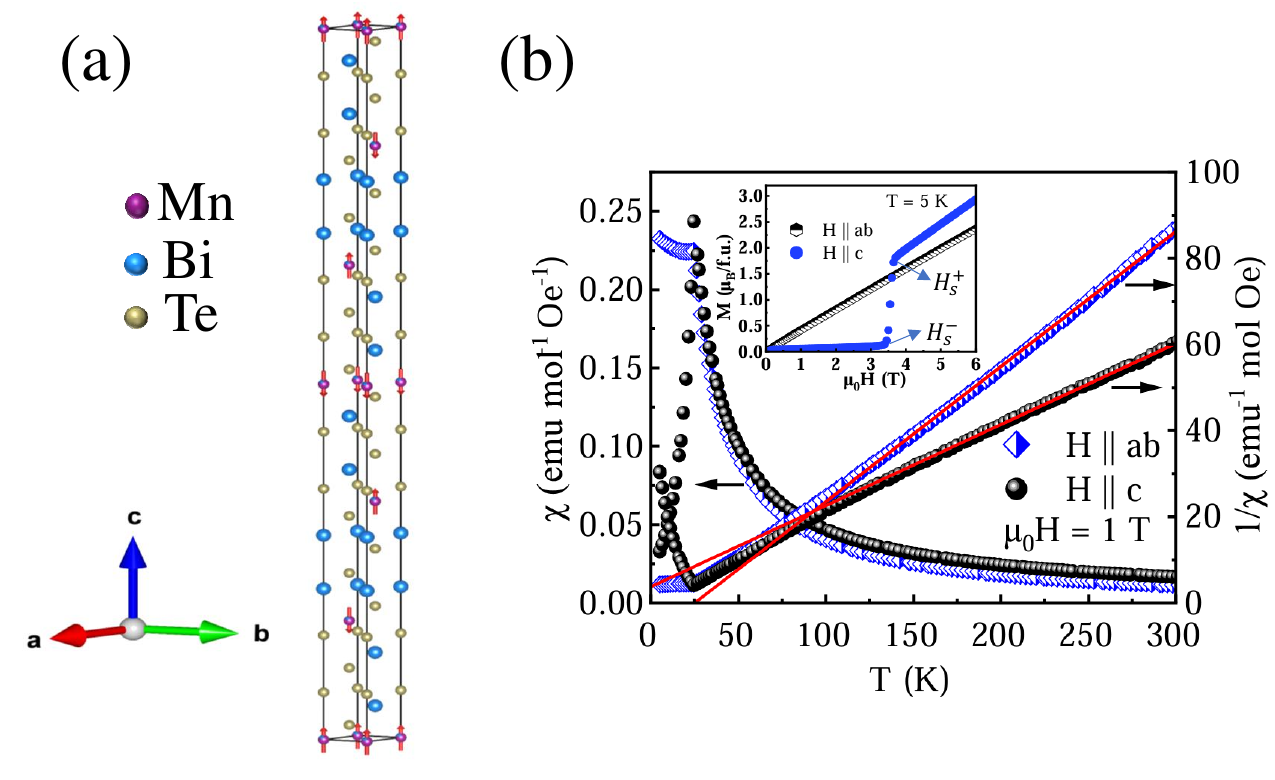}
\caption {(a) Unit cell of $\mathrm{MnBi_2Te_4}$ crystal. The red arrows indicate the spin orientation of each Mn atom.
        (b) Variation of zero-field cooled magnetization with the temperature at $ \mathrm{\mu_0H = 1T} $ applied along ab plane and c axis. The red line indicates the Curie-Weiss Fit for both the ab plane and the c axis. Inset: Field dependent magnetization for both $\mathrm{H || ab}  $ and $\mathrm{H || c}  $ at $\mathrm{T = 5 K.}$ $\mathrm{H_s^+}$ and $\mathrm{H_s^-}$ indicates the upper and lower boundary of the first critical field of the spin-flop transition along $\mathrm{H || c}$.}
\label{M1}
\end{figure}

The recently discovered MnBi$_2$Te$_4$ is an antiferromagnetic topological insulator (AFTI) with a gapped-out surface state. There are seven atomic layers in the sequence Te-Bi-Te-Mn-Te-Bi-Te making the septuple layers (SLs) which are stacked on each other along the z-axis \cite{PhysRevMaterials.5.124206, He2020, Otrokov2019}. The unit cell is rhombohedral with $\mathrm{R\Bar{3}m}$ space group symmetry, as shown in Fig-\ref{M1} (a). The magnetism in this TI comes from its Mn atoms. The Mn spins are ferromagnetically aligned within each SL along the z-axis but the spins between two adjacent SLs are coupled antiferromagnetically which makes the ground state of this material an A-type AFM state \cite{doi:10.1126/sciadv.aaw5685, Klimovskikh2020,Lee, Sass2020, LiP}. It exhibits AFM order and undergoes a metamagnetic spin-flop transition at intermediate fields \cite{wang2021intrinsic, Val’kov2024}. The nature of the magnetic interactions in AFM TIs are important factors that dictate access to the quantum topological states \cite{Sass, Qian2023}. For example, the topological Hall effect observed in MnBi$_2$Te$_4$ could be related to the non-collinearity of the magnetic phase in the spin flop regime \cite{Bac2022}. The quantization of the anomalous Hall effect occurs in the absence of an external magnetic field due to spontaneously broken TRS in the SLs \cite{LiQ, Trang2021, Zhao2021}. Inelastic neutron scattering (INS) measurements on MnBi$_2$Te$_4$ suggest the existence of competing magnetic interactions leading to frustration in the system and the consequent proximity to a variety of magnetic phases, potentially including topologically non-trivial spin textures such skyrmions \cite{Bing2020}. Recent first-principles calculations also suggest the possibility of magnetically tunable topological quantum phase transitions in MnBi$_2$Te$_4$ \cite{Jiaheng2019}.

Here we combine the analysis of magnetic susceptibility and electrical transport in the presence of magnetic field to construct a complex magnetic phase diagram of bulk single crystalline MnBi$_2$Te$_4$. 

{\it Experimental details:--} Bulk single crystals of MnBi$_2$Te$_4$ are grown using both flux\cite{PhysRevMaterials.3.064202} and chemical vapour transport (CVT) method \cite{YAN2022164327} (see Appendix \hyperref[appendixA]{A} for details). The verification of chemical composition and structural characterization is carried out using energy-dispersive X-ray spectroscopy (EDS) and X-ray diffraction (XRD), respectively (see Appendix \hyperref[appendixA]{A} for details). The magnetization measurements are performed using Quantum Design PPMS (Physical Property Measurement System) with magnetic field applied along in-plane (H $||$ ab) and out-of-plane (H $||$ c) directions. Measurements of $\mathrm{\rho_{xx}}$ and $\mathrm{\rho_{xy}}$ of the bulk crystal are carried out using the standard six-probe lock-in technique down to 0.3 K in the presence of a 12 T magnetic field in Cryogenic Ltd. (UK) VTI.

{\it Results and discussion:--} The magnetization data shows the crystal having an out-of-plane easy axis with AFM order. Both the in-plane and out-of-plane magnetic susceptibility data are plotted together as shown in Fig.~\ref{M1}(b) at an applied field of 1 T. The sharp peak around 24 K indicates the AFM ordering temperature ($\mathrm{T_N}$). The inverse magnetic susceptibility above $\mathrm{T_N} $ follows the Curie-Weiss law $\mathrm{ \chi = \frac{C}{(T+\theta_{CW})}}$. The calculated effective magnetic moment is 6.51 $\mathrm{ \mu_B/Mn} $ with curie-temperature $\mathrm{\theta_{CW} = - 20.1 K}$ along $\mathrm{H || c}$, which implies AFM interaction among the interlayers. Along $\mathrm{H || ab}$, the $\mathrm{\theta_{CW}}$ is 25.3 K which confirms the FM interaction among the intra-layer spins. The $\mathrm{M}$ vs $\mathrm{H}$ plot shown in the inset of Fig.~\ref{M1}(b) is taken at 5 K which shows a sharp metamagnetic spin-flop transition at a critical field $\mathrm{H_{c1}}\simeq 3.4$ T. However, the M-H curve along $\mathrm{H || ab}$ shows no such transition and it is linear throughout the field scan. We also perform temperature-dependent magnetization measurements at different fields applied along the c-axis (see Fig.~\ref{A5} of Appendix \hyperref[appendixB]{B} for details). As expected, with the increase of the magnetic field, the ordering temperature $\mathrm{T_N}$ starts decreasing and is eventually suppressed to zero at high magnetic field. 

\begin{figure}
\includegraphics[width=0.8\linewidth]{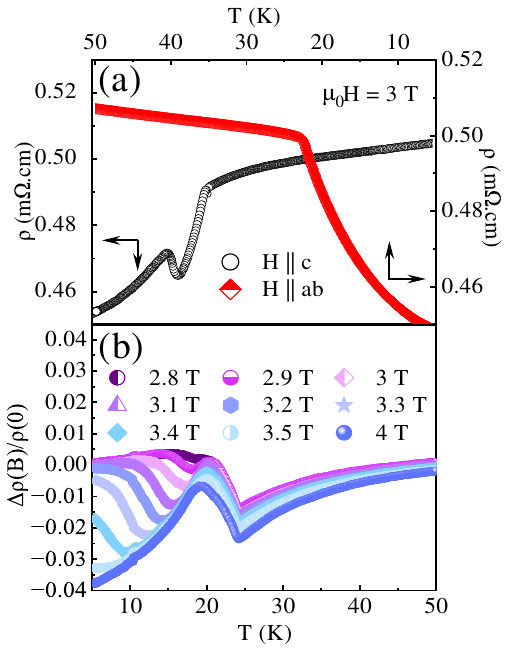}
\caption { Electrical resistivity variation with temperature measured at different fields along (a) $\mathrm{H || c} $ and $\mathrm{H || ab} $ and
        (b)  Magnetoresistance for different fields along  $\mathrm{H || c} $ .}
\label{M2}
\end{figure}

Fig.~\ref{M2}(a) shows the field-dependent electrical resistivity with the magnetic field of 3 T applied along $\mathrm{H || ab}$ and $ \mathrm{H || c} $. Along both directions, an abrupt change of slope in resistivity near the Neel temperature has been observed, consistent with earlier reports~\cite{PhysRevResearch.1.012011,Li}. More interestingly, the field-dependent resistivity for $ \mathrm{H || c} $ reveals the development of a distinct anomaly around 15 K in the intermediate magnetic field regime as shown in Fig.~\ref{M2}(a), something which has not been reported before, to the best of our knowledge. The anomalous behaviour in resistivity appears at a magnetic field strength of 2.9 T and is completely suppressed as the field is increased to 3.5 T (Fig.~ \ref{M2}(b)). 

\begin{figure}
\includegraphics[width= 0.8\linewidth]{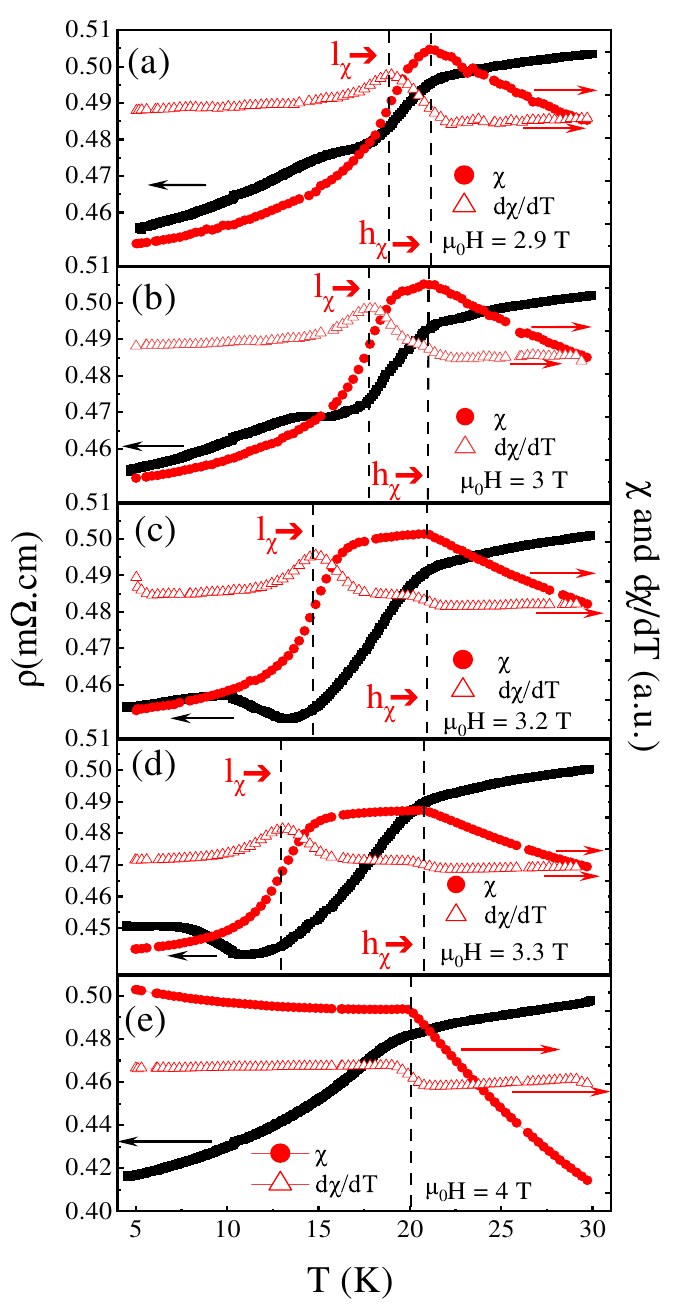}
\caption {Temperature dependence of electrical resistivity, susceptibility, and its derivative is plotted in the same graph for various constant fields along $\mathrm{H || c}$. The dashed line on the left tracks the anomaly in the derivative of susceptibility which vanishes in (e). The dashed line on the right of this figure points to the anomaly in susceptibility corresponding to $\mathrm{T_N}$. The anomalies in the susceptibility and its derivative have strong correspondence with the anomalies in the field-dependent resistivity.}
\label{M3}
\end{figure}

The combined plot of the temperature-dependent resistivity, magnetic susceptibility and its derivative in the presence of constant magnetic fields as shown in Fig.~\ref{M3} clearly suggests that the magnetic transitions and the resistive anomalies are correlated to each other. Tracking the evolution of the resistive anomaly as shown in Fig.~\ref{M3}(a) to \ref{M3}(d), and especially the correspondence between the peak in the 1st-order derivative of susceptibility and the anomaly in resistivity, it is clear that the anomaly is only observed within the metamagnetic regime. As the applied field is increased along $\mathrm{H || c}$, the plateau region of susceptibility gets broadened as well as the peak of its derivative and the anomalous dip in resistivity starts shifting towards lower temperature. Beyond the metamagnetic regime, the low-temperature peak in the derivative of susceptibility disappears along with the resistive anomaly (Fig.~\ref{M3}(e)). Contrary to the usual first-order metamagnetic transition \cite{SHANG}, the resistivity in the metamagnetic regime shows the absence of thermal hysteresis as in \cite{chattopadhyay2010disorder}, with the cooling and heating curves overlapping completely (see Fig.~\ref{A2}(c) in Appendix \hyperref[appendixC]{C} for details). As the magnetic field is applied, the resistance below 40 K starts decreasing and gives rise to negative magnetoresistance (MR). The enhancement in MR with the increasing magnetic fields signifies a field-induced phase transition as in Fig.~\ref{M2}(b).

In order to further investigate the nature of the magnetic phase transition, we utilize the magnetocaloric effect (MCE) to calculate the change of entropy $\mathrm{-\Delta S_M}$~\cite{Law2018}. The magnetic entropy change ($\mathrm{-\Delta S_M}$) is determined by Maxwell's thermodynamic relation using the isothermal magnetization curves within the temperature range 15 K to 34 K at regular intervals of 1 K (see  Fig~\ref{A3}(a) in Appendix \hyperref[appendixB]{B} for details)
\begin{eqnarray}
\mathrm{\Delta S_M(T,H)}&=&\mathrm{S_M(T,H)-S_M(T,0)} \\
&=&\mathrm{\int_{0}^{H}\left(\frac{\partial M(T,H)}{\partial T}\right)dH}
\label{eqn}
\end{eqnarray}
\begin{figure}
\includegraphics[width=0.7\linewidth]{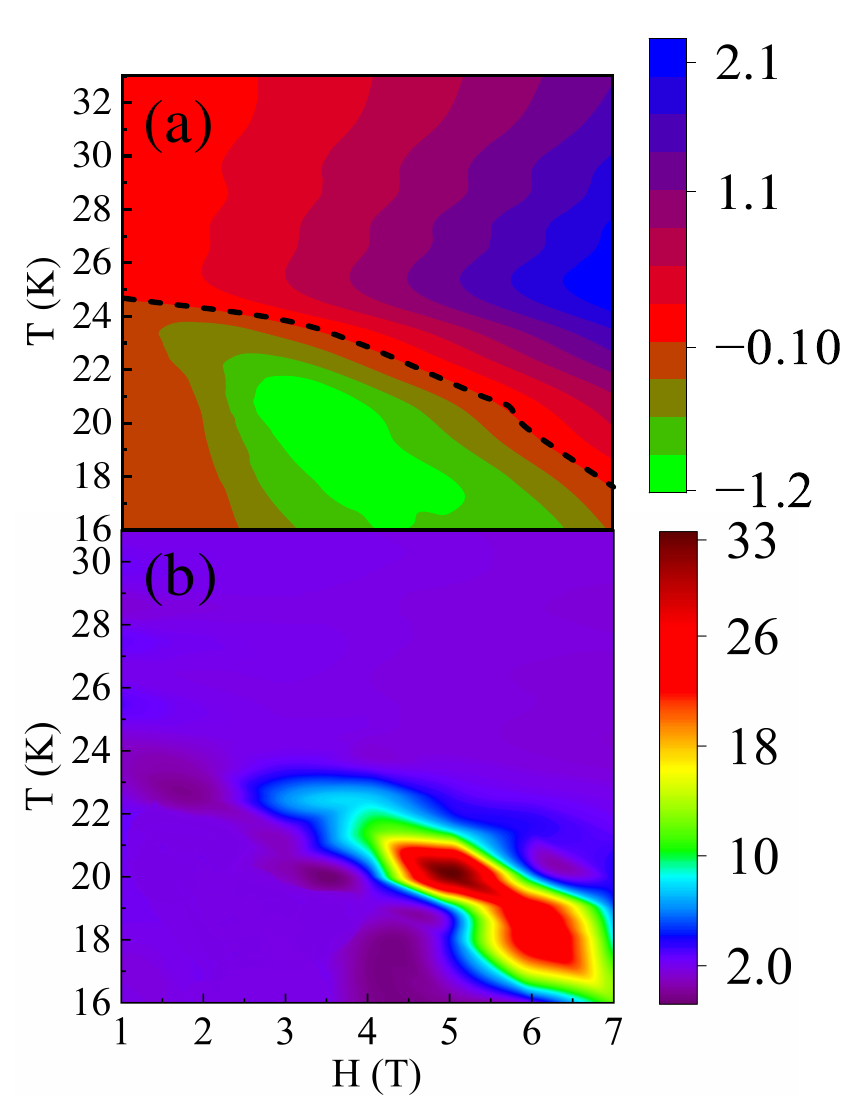}
\caption {(a) 2d projection of three-dimensional curve of entropy ($\mathrm{\Delta S_M(T,H)}$) as a function of temperature and magnetic field.
        (b) Field and temperature dependencies of exponent n. }
\label{M4}
\end{figure}
\begin{figure*}
\includegraphics[width=1\linewidth]{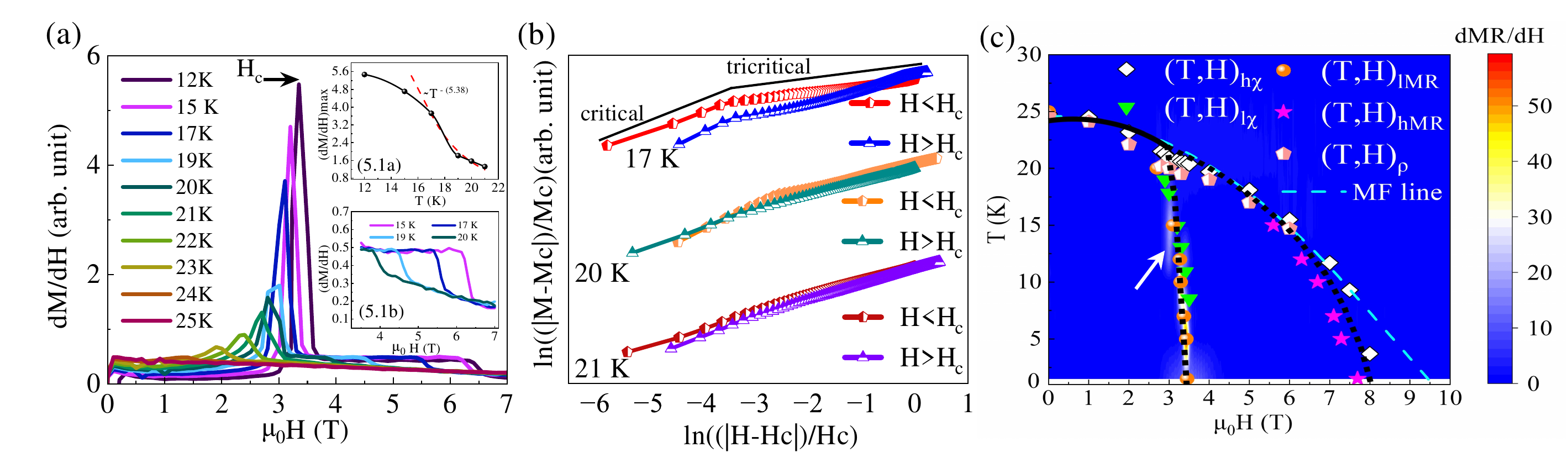}
\caption {(a) Field derivative of magnetization as a function of magnetic field. The sharp peaks refer to the MMT. $\mathrm{H_c (\simeq H_{c1})}$ refers to the field where $\mathrm{\frac{dM}{dH}}$ is maximum. Inset: (5.1a) Temperature dependence of the maxima of differential susceptibility. The black circles correspond to low-field maxima. The black line through the circles is a guide to the eye, whereas the red-dashed line indicates the power law dependency of $\mathrm{T^{-n}}$.     
(5.1b) $\mathrm{\frac{dM}{dH}}$ from 4 T to 7 T shows a sudden drop which is related to $\mathrm{H_{c2}}$.
(b) logarithmic plots of normalized magnetization and normalized field for different isotherms.
(c) Magnetic phase diagram with applied field $|| $ c axis presented an H-T colour plot with $\mathrm{\frac{dMR}{dH}}$(from Fig.~\ref{A6}(a)) represented as the colour
scale. The white patches of $\mathrm{\frac{dMR}{dH}}$ (indicated by the white arrow) correspond to the first spin-flop transition: $\mathrm{(T, H)_{lMR}}$ which exactly matches with the $\mathrm{(T, H)_{l\chi}}$. As indicated, the scattered points correspond to different temperatures obtained from different magnetization and transport measurements. The solid and dashed lines guide the eye for different regimes as written in the text.
 }
\label{M5}
\end{figure*}
With the increasing field the value of $\mathrm{\Delta S_M(T, H)}$ changes from negative to positive as the interaction between inter-layered spins becomes FM from AFM in the spin flop regime. For $\mathrm{\mu_0H} = 7$ T, the entropy change $\mathrm{\Delta S_M(T,H)}$ exhibits a maximum value of 2.27 J Kg$^{-1}$ K$^{-1}$ around the AFM transition temperature. The two-dimensional projection of the field and temperature-dependent magnetic entropy is plotted in Fig.~\ref{M4}(a). The nature of phase transition can in principle be determined from the shape of a three-dimensional MCE graph \cite{Law2018}. It is clear from Fig.~\ref{M4}(a) that there exists a phase boundary (shown by the dashed line in Fig.~\ref{M4}(a)) separating the negative and positive $\mathrm{\Delta S_M(T, H)}$ just below $\mathrm{T_N}$ within a narrow temperature window in the intermediate magnetic field regime. We further utilise a quantitative approach for determining the phase transition using the dependency of the magnetocaloric entropy change on magnetic field. The change in magnetic entropy is linearly proportional to the $\mathrm{n}$th power of the magnetic field, $\mathrm{\Delta S_M(T,H) \propto H^n}$, where the `local' power law exponent $\mathrm{n}$ is determined using the relation 
\begin{equation*}
    \mathrm{n(T,H)}= \mathrm{\frac{dln|\Delta S_M|}{dlnH}}
\end{equation*}
If the value of $\mathrm{n}$ lies between $0 \rightarrow{2}$ then the nature of phase transition is 2nd order. For 1st order phase transition, $\mathrm{n}$ always takes values $\mathrm{n}>2$.
We observe that the value of $\mathrm{n}$ overshoots 2 across the phase boundary (Fig.~\ref{M4}(b)). At $\mathrm{T<T_N}$ as the magnetic field is increased the magnitude of the n reaches its maximum suggesting a field-induced first-order phase transition. In order to understand the phase diagram of the system in greater detail, we also measure $\mathrm{\rho_{xx}}$ and $\mathrm{\rho_{xy}}$ as a function of the magnetic field along the c-axis (see Fig.~\ref{A6} in appendix-\hyperref[appendixC]{C} for more details).

One more important observation that demands further attention is the field dependency of differential susceptibility ($\mathrm{\frac{dM}{dH}}$) as shown in Fig.~\ref{M5}(a). The maxima in $\mathrm{\frac {dM}{dH}} $ vs H plot corresponds to the lower critical field ($\mathrm{H_{c1}\simeq H_{c}}$) of the MMT. The maximum in the differential susceptibility, $\mathrm{(\frac {dM}{dH})_{max}} $, corresponding to the lower critical field, increases in amplitude and becomes increasingly sharper as the temperature is lowered. On the other hand, the sharp drop beyond the plateau region corresponding to the higher critical field increases with the lowering of temperature. Moreover, the $\mathrm{(\frac {dM}{dH})_{max}} $ diverges following the power law : $\mathrm{(\frac {dM}{dH})_{max}  \sim T^{-n}}$ (with n= 5.38) in the vicinity of $\mathrm{T_N}$. At lower temperature, $\mathrm{(\frac {dM}{dH})_{max}} $ tends to saturate. Such behaviour is typical of the long-wavelength quantum critical fluctuations over the H–T phase diagram in the vicinity of the critical field and has been reported earlier in rare-earth systems \cite{15Das}. Fig~\ref{M5}(b) indicates the existence of tricriticality in MnBi$_2$Te$_4$ (see Appendix \hyperref[appendixD]{D} for details). Two different slopes of normalized magnetization have been observed below the tricritical temperature ($\mathrm{T_t = 21.2 K}$, see appendix-\hyperref[appendixD]{D}), which disappear as the tricritical point (TCP) is reached. The disappearance of two distinct slopes near the TCP is the result of the interplay between the second-order phase and tricritical phase present in the system.

The overall picture that emerges from the magnetization and transport properties is summarized by the phase diagram in Fig.~\ref{M5}(c). The phase diagram is constructed as follows. The location of the two maxima of the first derivative of MR (obtained from Fig.~\ref{A6}(a)) has been taken first. We label the $\mathrm{(T, H)}$ co-ordinates corresponding to the low-temperature anomalies in the temperature derivative of susceptibilities (plotted in Fig.~\ref{M3}) as $\mathrm{(T, H)_{l\chi}}$ and $\mathrm{(T, H)_{h\chi}}$, respectively and plot them separately Fig.~\ref{M5}(c). These points overlap with those obtained from low field and high field anomalies in the MR data (plotted in Fig.~\ref{A6}(a)) labelled as $\mathrm{(T, H)_{l, MR}}$ and $\mathrm{(T, H)_{h, MR}}$, respectively in Fig.~\ref{M5}(c). Finally, we incorporate the $\mathrm{(T, H)}$ co-ordinates, labelled as $\mathrm{(T, H)_{\rho}}$, obtained from the maximum in the temperature derivative of resistivity in the intermediate field regime into Fig.~\ref{M5}(c). 

The major highlight of the phase diagram is the existence of a TCP at $\mathrm{T = 21.2}$ K and at $\mathrm{B} = 2.8$ T, which distinguishes AFM, PM and the intermediate canted AFM (c-AFM) state of MMT. At low fields, $\mathrm{T_N(B)}$ follows a mean-field-like power law (marked as the blue dashed line in fig-\ref{M5}(c)):  $\mathrm{T_N(B) = T_N(0)[1-(\frac{B}{B_0})^2}]$, starting from its zero-field value $\mathrm{T_N(0)=24.6}$ K. Extrapolation of the mean-field curve suggests that a conventional QCP would have been obtained at $\mathrm{H_c=9.5}$ T. However, there is a clear deviation from the mean-field behaviour at intermediate fields when the metamagnetic phase starts emerging after crossing the TCP, suggesting that the nature of AFM transition changes from 2nd order to 1st order at TCP. Starting from TCP, as the magnetic field is increased further, the 2nd order phase line (the solid black line in Fig.~\ref{M5}(c)) bifurcates into two separate lines of first-order transition (two black dotted lines in Fig.~\ref{M5}(c)), which terminate at quantum critical endpoints (QCEPs).

{\it Conclusion:--} We have combined the magnetic and transport properties of bulk single-crystal of MnBi$_2$Te$_4$ to construct a complex phase diagram that exhibits a TCP with coexisting AFM phase, an intermediate mixed magnetic phase and FM phase. The second-order AFM-PM transition line splits at the TCP into two separate first-order lines which terminate at QCEPs. Recently, based on magnetic susceptibility measurements, a metamagnetic tricritical point quite similar to the present case has been reported in MnBi$_4$Te$_7$~\cite{HuiZhang}. However, there are important differences as well: MnBi$_4$Te$_7$ has much lower $\mathrm{T_N \sim 13}$ K. Moreover, in contrast to MnBi$_2$Te$_4$, as described in the present article, the tricritical point in MnBi$_4$Te$_7$ was observed at a very weak magnetic field and most importantly, without any evidence of quantum critical endpoints. In particular, we observe a magnetic field-induced `intermediate' quantum phase spanning a large area in the ($\mathrm{H, T}$) plane which leaves its imprint on the electrical transport properties along the c-axis.

{\it Acknowledgements:--} The authors acknowledge IIT Kanpur and the Department of Science and Technology, India, [Order No.
DST/NM/TUE/QM-06/2019 (G)] for financial support. T.D. thanks PMRF for financial support. 

\section{Appendix A: Growth and Characterization}\label{appendixA}
\subsection{1. Single crystal Growth}
MBT single crystals are grown using the chemical vapour transport (CVT) method with iodine as the transport agent. The initial precursors, Mn (99.99 \%,Alfa Aesar), Bi (99.999 \%,
Alfa Aesar), and Te (99.99\%, Alfa Aesar) powders, are mixed
in a molecular ratio of 2:2:5. The pelletized homogeneous mixture, along with $\mathrm{I_2}$, was then enclosed in a
vacuum-sealed quartz ampule. The growth process utilizes a
TG3-1200 gradient tube furnace, with the reaction zone (hot
end) maintained at $585^{\circ}$C and the growth zone (cold end)
at $575^{\circ}$C. Both zones are heated at a rate of $1^{\circ}$C/m. After 15 days of dwelling in the furnace, the ampule is cooled to room temperature, resulting in the
collection of shiny crystals with an average length of approximately
2 mm. All the data was collected using CVT-grown crystals. However, the self-flux method \cite{PhysRevMaterials.3.064202} has also been used to make another single crystal to investigate whether the antisite defect has any reciprocity with the resistive anomaly. 
\begin{figure}[htp]
\includegraphics[width=0.7\linewidth]{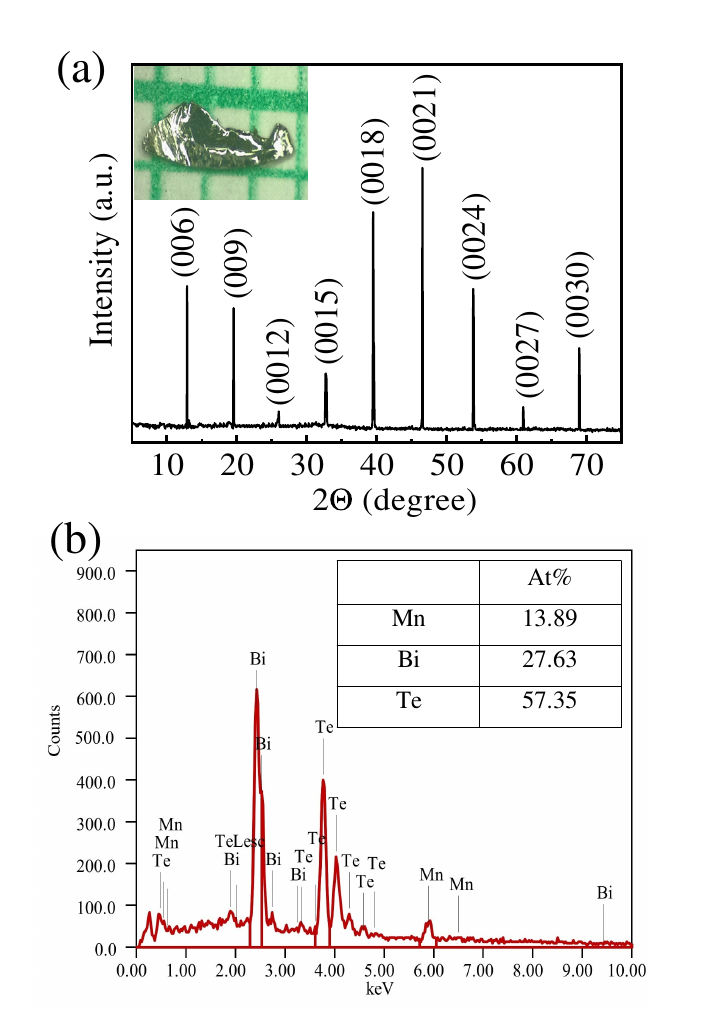}
\caption {(a) XRD $\mathrm{\theta-2\theta}$ scan of the MnBi$_2$Te$_4$ single crystal. Inset: Optical image of the as-grown crystal on a mm grid.
(b) EDS spectrum from a freshly
cleaved crystal surface showing the presence of Mn, Bi, and Te peaks
with nominal stoichiometric ratio (inset).}
\label{A1}
\end{figure}

\begin{figure}
\includegraphics[width=.7\linewidth]{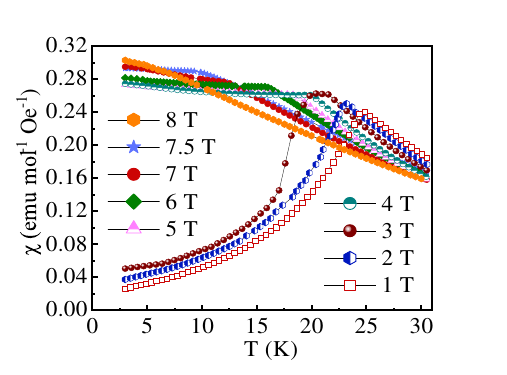}
\caption { The MT curve at different fields along $\mathrm{H || c}$ }.
\label{A5}
\end{figure}
\begin{figure}
\includegraphics[width=.7\linewidth]{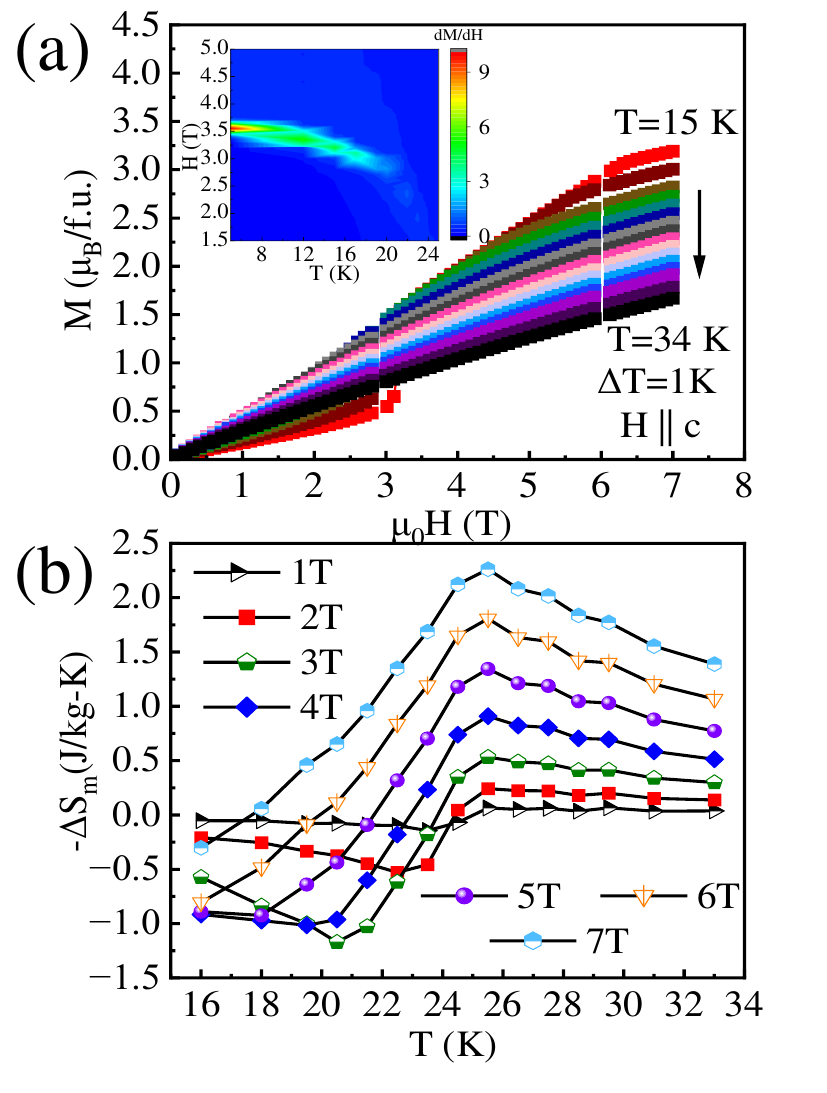}
\caption {(a) Isothermal magnetization curves at different temperatures. [Inset: The first-order derivative of magnetization $\mathrm{\frac{dM}{dH}}$ has been plotted as a function of the applied field and temperature. The MMT region becomes narrower approaching the TCP]. 
(b) Magnetic entropy change ($\mathrm{-\Delta S_M}$) as a function of temperature for different magnetic field changes up to $\mathrm{7}$ T.} 
\label{A3}
\end{figure}
\begin{figure}[htp]
\includegraphics[width=1\linewidth]{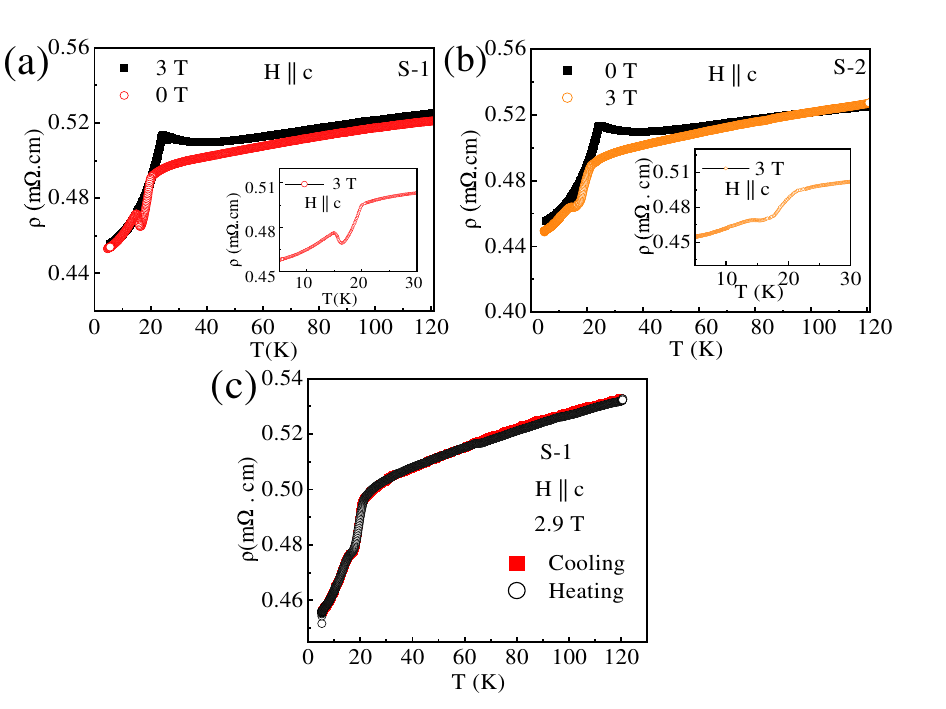}
\caption {Field-dependent resistivity of (a) CVT-grown sample (S-1) (b) self flux-grown sample (S-2). (c) Resistivity data of S-1 taken at both heating and cooling cycle at $\mathrm{2.9}$ T. }
\label{A2}
\end{figure}
\begin{figure}
\includegraphics[width=0.7\linewidth]{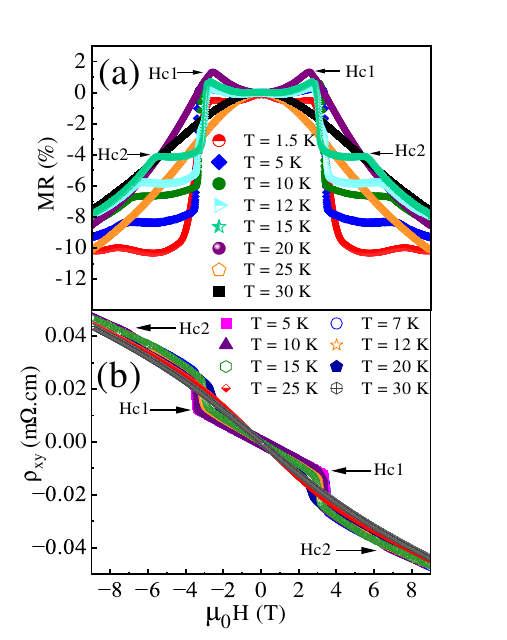}
\caption {(a) The MR curve along $\mathrm{H || c}$ at different temperatures.
        (b) The Hall resistivity along $\mathrm{H || c}$ at different temperatures. }
\label{A6}
\end{figure}
\begin{figure}
\includegraphics[width=1\linewidth]{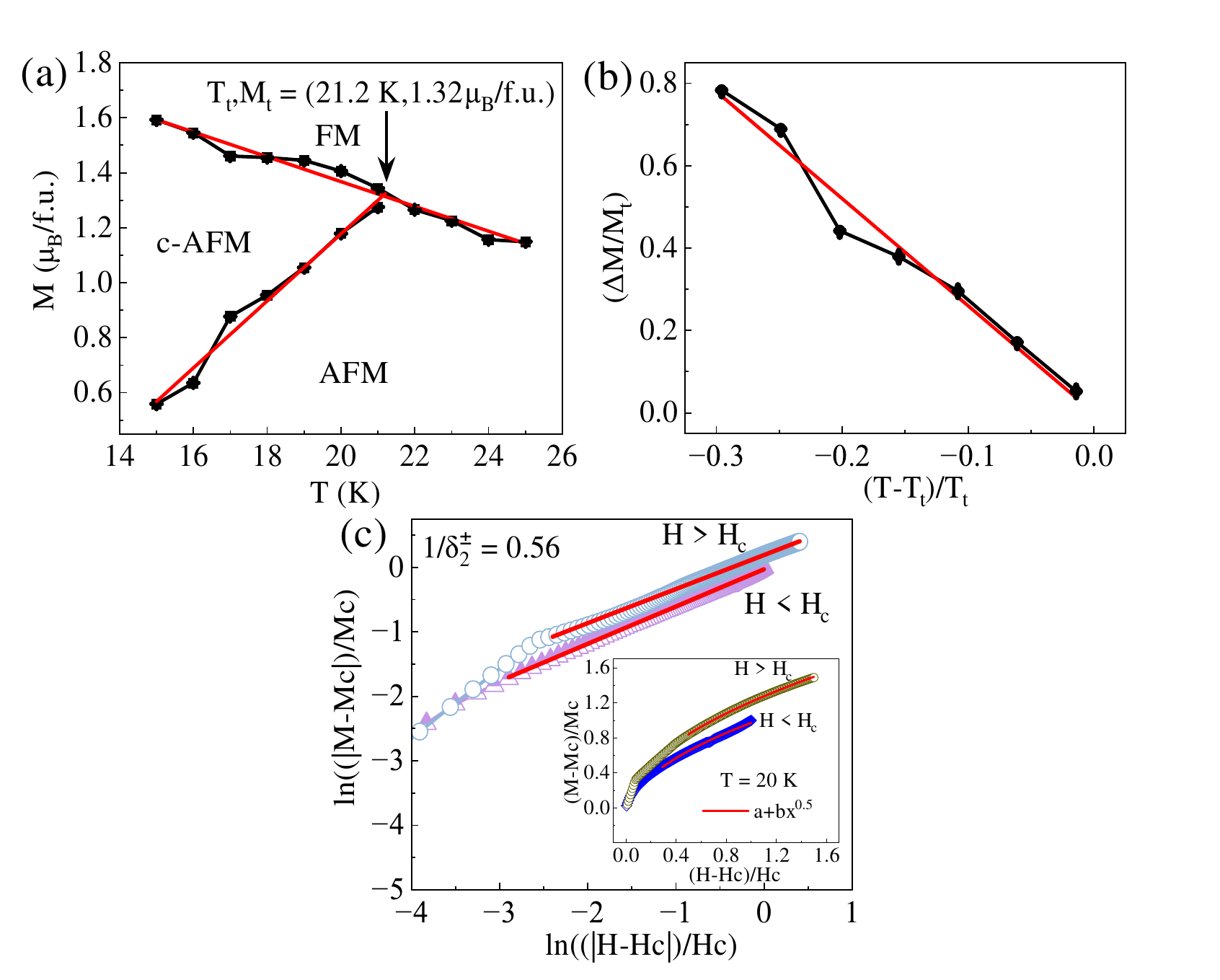}
\caption {(a) The M-T phase diagram is plotted by taking the magnetizations corresponding to the upper and lower value ($\mathrm{H_s^+, H_s^-}$ as shown in the inset of fig~\ref{M1}(b)) of the first critical field $\mathrm{H_{c1}}$. $\mathrm{T_t, M_t}$ is the tricritical temperature and tricritical magnetization respectively.  
(b) Normalized magnetization has been plotted to show the discontinuity near TCP. The red line corresponds to $\mathrm{\beta_2 ^\pm \approx 1}$. 
(c) log-log plot of normalized magnetization and normalized field at 20 K. Tricritical exponent $\mathrm{\delta_2^\pm}$ has been obtained using the relation $\mathrm{M \sim H^\frac{1}{\delta_2}}$. The red line indicates the linear slopes.
Inset: Normalized magnetization as a function of normalized magnetic field at $\mathrm{T=20 K (< T_t)}.$ The red line is fitted using the relation $\mathrm{y=a+bx^{0.5}}.$}
\label{A7}
\end{figure} 
\subsection{2. Characterization}
Energy dispersion X-ray spectroscopy(EDS) and X-ray diffraction(XRD) measurements have been carried out to determine the elemental compositions and crystallinity of bulk-MBT, respectively. 
The XRD spectra are shown in Fig.~\ref{A1}(a) with peaks along (0 0 l) direction, indicating the high quality of the as-grown crystal with the crystallographic c axis being perpendicular to the crystal ab plane. Fig.~\ref{A1}(b) presents the EDS spectra with prominent Mn, Bi, and Te peaks obtained from a freshly cleaved surface of an MBT crystal. From the EDS measurements performed across various crystal surfaces, we get an average stoichiometric ratio of Mn, Bi, and Te to be 1.00:2.00:4.15, which is close to the nominal stoichiometric ratio for MBT. This also confirms that there is no Mn deficiency present in the measured crystal.

\section{Appendix B: Magnetic Measurements}\label{appendixB}
{\subsection{1. Susceptibility}}
Fig.~\ref{A5} shows $\mathrm{\chi(T)}$ at different magnetic fields along $\mathrm{H || c}$. As the magnetic field is increased, the transition temperature $\mathrm{T_N}$ starts decreasing and becomes undetectable at an applied field of $\mathrm{8}$ T. This behaviour indicates that the antiferromagnetic order is suppressed to $\mathrm{T_N \rightarrow{0}}$ by applying a field of 8 T along c axis.

{\subsection{2. Magnetocaloric Effect}}
Taking into account the direct correlation between the magnetic entropy change ($\mathrm{(-\Delta S_M)}$) and the local exponent (n), ($\mathrm{(-\Delta S_M)}$) has been calculated using the isothermal magnetization. Fig.~\ref{A3}(a) shows the isothermal magnetization as a function of magnetic field around $\mathrm{T_N}$. The $\mathrm{(-\Delta S_M)}$ curves are obtained from the M(H) curves using Maxwell’s relation as discussed earlier in Eq.-\ref{eqn}. As shown in Fig.~\ref{A3}(b) for $\mathrm{H || c}$, a crossover to inverse magnetic entropy change [negative ($\mathrm{(-\Delta S_M)}$)], occurs below $\mathrm{T_N}$ which consistent with out-of-plane AFM transition. The maximum value of magnetic entropy change ($\mathrm{(-\Delta S_M)}$) at 7 T reaches 2.26 J/(kg.K) along $\mathrm{H || c}$.

\section{Appendix C: Magneto-transport Measurements}\label{appendixC}
{\subsection{1. Comparison of Resistivity Data}}
Antisite defects such as $\mathrm{Mn_{Bi}}$ in the Bi layer of $\mathrm{MnBi_{2}Te_{4}}$ can affect the magnetic and electronic structure of this material greatly \cite{Wu2023, Huang, Garnica2022, Yang_2025}. So, a comparative study of the CVT-grown sample (S-1) and the flux-grown sample (S-2) has been carried out. The zero-field resistivity of S-1 and S-2 have been plotted in Fig.~\ref{A2}, both showing resistive anomaly in the intermediate field regime. The anomaly is more distinguishable for the Mn-rich S-1 sample as compared to S-2 (Mn:Bi:Te = 0.977:2.268:4) at the same applied field. This observation clarifies that the spin-flop field range along with the resistive anomaly may vary from sample to sample because of the site mixing effect but the anomaly in resistivity is arising only due to metamagnetism. The resistivity data has been taken in both the cooling and heating cycle (the measurement of heating was made after the sample was cooled in zero field) to observe the possible hysteresis of a metamagnet \cite{LIU201526}. No hysteretic behaviour has been observed, as shown in Fig.~\ref{A2}(c).

{\subsection{2. Magnetoresistance}}
The magnetoresistance (MR) as well as the Hall resistivity are shown in Fig-\ref{A6}. We observe a sharp spin flop transition occurring at critical fields $\mathrm{H_{c1}}$ below the $\mathrm{T_N}$. As the magnetic field is increased further and reaches a higher critical field $\mathrm{H_{c2}}$, the canted-AFM spins in the metamagnetic regime polarize completely. The magnetic field dependence of Hall resistivity also undergoes distinct changes of slope at the two critical fields $\mathrm{H_{c1}}$ and $\mathrm{H_{c2}}$ which vanish above $\mathrm{T_N}$.

\section{Appendix D: Tricritical Analysis}\label{appendixD} 

The Ising model of the metamagnet indicates the existence of a tricritical point in the magnetic phase diagram when the competing interaction ratio $\mathrm{\lambda = \frac{J_{||}}{J_{\perp}}< -\frac{3}{5}}$ \cite{kincaid}. For $\mathrm{MnBi_2Te_4}$, the value of $\mathrm{\lambda = -4.09 <\frac{3}{5} }$ \cite{Otrokov2019}, which confirms the emergence of a tricritical point in this material. According to the mean-field theory, multiple phases approach the tricritical point linearly with $\mathrm{\beta_2^+}\approx 1$ and $\mathrm{\beta_2^-}\approx 1$ where $\mathrm{\beta_2}$ is one of the mean-field exponents. Fig~\ref{A7} (a) shows that the AFM, spin-flop and FM phases meet the TCP linearly at 21.2 K. Along with it, the discontinuity in the nonordering magnetization should follow the linear law $\mathrm{\frac{\Delta M}{M_t} = A(1-\frac{T}{T_t})^{\beta_2}}$ as shown in Fig~\ref{A7}(b) and in this case the obtained values are: $\mathrm{A= 2.947 \pm 0.47, \beta_2 = 1.11 \pm 0.11, T_t = 21.2 \pm .05 }$K. Another tricritical exponent $\mathrm{\delta^\pm_2}=0.56 \pm 0.08$ has been obtained by plotting the log-log normalized magnetization against the normalized magnetic field at 20K($\mathrm{<T_t}$) as shown in Fig.~\ref{A7}(c). At this temperature, for both $\mathrm{H < H_c}$ and $\mathrm{H > H_c}$, two distinct slopes (one away from the TCP has slope $\mathrm{\approx 0.56}$ and another close to the TCP has slope $\mathrm{> 1}$ ) have been observed where $\mathrm{H_c (\simeq H_{c1})}$ is the magnetic field at which $\mathrm{\frac{dM}{dH}}$ is the maximum (shown in Fig~\ref{M5}(a)) and $\mathrm{M_c}$ is the magnetization corresponding to $\mathrm{H_c}$. This distinction disappears as the TCP is reached as shown in Fig~\ref{M5}(c).

\end{document}